\input harvmac.tex
\input epsf






\Title{}{Strings and Discrete Fluxes of QCD.}

\centerline{$\quad$ { Z. Guralnik}}
\smallskip
\centerline{{\sl University of Pennsylvania}}
\centerline{{\sl Philadelphia PA, 19104}}
\centerline{{\tt guralnik@ovrut.hep.upenn.edu}}

\vskip .3in

We study discrete fluxes in four dimensional $SU(N)$ gauge theories 
with a mass gap by using brane compactifications which give 
${\cal{N}} = 1$ or ${\cal{N}} = 0$ supersymmetry.
We show that when such theories are compactified further on
a torus,  the t'Hooft magnetic flux $m$ is related to the NS two-form 
modulus $B$ by $B = 2\pi {m\over N}$.  
These values of $B$ label degenerate brane vacua,  giving
a simple demonstration of magnetic screening.
Furthermore, for these values of $B$ one has a conventional gauge theory 
on a commutative torus,  without having to perform any T-dualities. 
Because of the mass gap, a generic $B$ does not give a 
four dimensional
gauge theory on a non-commutative torus.  The Kaluza-Klein modes which
must be integrated out to give a four dimensional theory 
decouple only when $B=2\pi {m\over N}$. 
Finally we show that $2\pi {m\over N}$ behaves like a two form modulus of
the QCD string.  This confirms a previous conjecture based on properties of
large $N$ QCD suggesting a T-duality invariance.


\Date{}


\lref\mqcdwitten{E. Witten, {\it ``Branes and the dynamics of QCD,''}
		Nucl. Phys. {\bf B507} (1997) 658-690,  hep-th/9706109.}
\lref\mqcdooguri{K. Hori and H. Ooguri, {\it ``Strong coupling dynamics 
		of four-dimensional N=1 gauge 
		theories from M theory five-brane.''} 
		Adv. Theor. Math. Phys. {\bf 1} (1998) 1,  
		hep-th/9706082.}
\lref\mqcdstrassler{A. Hanany,  M. Strassler and A. Zaffaroni, 
		{\it ``Confinement and strings in MQCD,''}
			Nucl. Phys. {\bf B513} (1998) 87, hep-th/9707244.}
\lref\maldads{J. Maldacena, {\it ``The large N limit of superconformal 
		field theories and supergravity,''} 
		Adv. Theor. Math. Phys. {\bf 2} (1998) 231, 
		hep-th/9711200.}
\lref\adskleb{S. Gubser, I. Klebanov, and A. Polyakov, {\it ``Gauge theory
		correlators from noncritical string theory,''}
		Phys. Lett. {\bf B428} (1998) 105-114, hep-th/9802109.} 
\lref\ads{E. Witten, {\it ``Anti-de-Sitter space, thermal phase
		transition,  and confinement in gauge theories,''}
		Adv. Theor. Math. Phys. {\bf 2} (1998) 505,  
		hep-th/9803131. }
\lref\wittenads{E. Witten, {\it ``Anti-de-Sitter space and holography,''}
		Adv. Theor. Math. Phys. {\bf 2} (1998) 253,  
		hep-th/9802150.} 
\lref\edflux{E. Witten, {\it ``ADS/CFT correspondence and topological 
		field theory,''}
		JHEP {\bf 9812}  (1998) 012, hep-th/9812012.}
\lref\douglas{M. Douglas, {\it ``Conformal field theory techniques 
		in large N Yang-Mills theory,''}
		Cargese Workshop on Strings, Conformal Models 
		and Topological Field Theories, 
		Cargese, France, May 12-26, 1993,  hep-th/9311130.}
\lref\meone{Z. Guralnik, {\it ``T-duality of large N QCD,''} 
		1998 Paris workshop on Quantum Chromodynamics.
		hep-th/9903021.} 
\lref\metwo{Z. Guralnik, {\it ``Duality of large N Yang-Mills theory 
		on $T^2 \times R^n$,''} hep-th/9804057.}
\lref\edsolns{E. Witten, {\it ``Solutions of four-dimensional field 
		theories via M Theory,}
		Nucl. Phys. {\bf B500} (1997) 3, hep-th/9703166.}
\lref\zram{Z. Guralnik and S. Ramgoolam, {\it ``Torons and D-brane 
		bound states,''}
		Nucl. Phys. {\bf B499} (1997) 241, hep-th/9702099.}
\lref\noncomm{ A. Connes, M. Douglas, A. Schwarz, 
	{\it "Noncommutative geometry and Matrix theory:
	 compactification on tori,"} JHEP {\bf 9802} (1998) 003, 
	hep-th/9711162.}
\lref\hulldoug{ M. Douglas and C. Hull, {\it "D-branes and
		the noncommutative torus,"} JHEP {\bf 9802} (1998) 008, 
		hep-th/9711165.}
\lref\jabone{F. Ardalan, H. Arfaei and M.M. Sheikh-Jabbari, 
	{\it ``Mixed branes and M(atrix) theory on non-commutative torus,''}
	hep-th/9803067.}
\lref\jabtwo{F. Ardalan, H. Arfaei and M.M. Sheikh-Jabbari,
	{\it ``Noncommutative geometry from strings and branes,''}
	JHEP {\bf 9902} (1999) 016,
	hep-th/9810072.}  
\lref\peimingho{C.-S Chu and P.-M Ho, {\it ``Noncommutative open 
		string and D-brane,''} Nucl. Phys. {\bf B550} (1999) 151, 
		hep-th/9812219.}
\lref\thooftor{G. 'tHooft, {\it ``Some twisted self-dual solutions for 
		the Yang-Mills 
		equations on a hypertorus,''} 
		Comm. Math. Phys {\bf 81} (1981) 267.}
\lref\brodie{J. Brodie, {\it ``Fractional branes, confinement, and 
		dynamically generated 
		superpotentials,''}
		Nucl. Phys. {\bf B532} (1998) 137, hep-th/9803140.}
\lref\dougbranes{ M. Douglas, {\it "Branes within branes,"}
		Cargese 1997, Strings, branes and dualities 267, 
               hep-th/9512077.}
\lref\thooft{ G. 't Hooft, {\it "A property of electric and
		magnetic flux in non-abelian gauge theories,"}
	       Nucl. Phys. {\bf B153} (1979) 141.}
\lref\Aharonyberk{ O. Aharony, M. Berkooz, N. Seiberg, 
	{\it ``Light cone description of (2,0) superconformal theories in
	six-dimensions,''} Adv. Theor. Math. Phys. {\bf 2} (1998) 119,
	hep-th/9712117. }
\lref\nekrasov{A. Astashkevich, N. Nekrasov, A. Schwarz, 
	{\it ``Instantons on noncommutative $R^4$ and (2,0) superconformal 
	six-dimensional theory,''} Commun. Math. Phys. {\bf 198} (1998) 689,
        hep-th/9810147. }


\newsec{ Introduction } 

Recently, an improved understanding of the infrared properties of QCD 
has been gained by studying classical configurations of M theory.  
In \mqcdwitten\mqcdstrassler\mqcdooguri,  a theory in the same 
universality class as 
four dimensional ${\cal N} =1$ QCD, known as MQCD, was 
studied by wrapping an M theory 
fivebrane on a
holomorphic curve  $\Sigma$ embedded in the spacetime 
$R^{10} \times S^1$.  
${\cal{N}} =0$ MQCD was also constructed in \mqcdwitten\ by taking
$\Sigma$ to be a non-holomorphic two-cycle of minimal area.
The results of this paper are applicable to either case.
We will further compactify the theory on a torus of finite size,  
so that the M-fivebrane 
wraps $\Sigma \times T^2$ which is embedded in 
$R^8 \times T^2 \times S^1$.
In the absence of fundamental matter, $SU(N)$ QCD  
on a two-torus 
has a discrete magnetic flux $m$ which is defined modulo $N$ \thooft.
The supergravity quantities
corresponding to discrete Yang-Mills fluxes in the ADS/CFT approach 
\adskleb\maldads\wittenads\ads\ were discussed
in \edflux.   
In that context, the discrete fluxes appeared as states in a topological
field theory.
In the present context, we will see that the 't Hooft flux 
corresponds to the background M-theory three-form through the relation
\eqn\reln{\int_{T^2\times S^1} C = 2\pi {m\over N}.} 
Since this result is independent of the radius of the $S^1$, we expect it
to persist in the IIA limit,  giving 
\eqn\IIAlimit{ B = \int_{T^2} B_{NS} = 2\pi {m\over N}.} 

This relation is peculiar to the case in which there is a mass gap.  
The background three-form is of course not really discretely quantized.
The vacuum energy of the five-brane has a periodic dependance
on this background,  with minima at $\int C = 2\pi {m\over N}$.  
At these minima,  there exists a limit in which the low energy theory 
is QCD,  with t' Hooft flux $m$.  For other values of $\int C$,  one 
might naively expect
to find a gauge theory on a non-commutative torus, \noncomm\hulldoug\
but this is not the case.
We will show that for $\int C$ away from the minima,  it is impossible 
to decouple the Kaluza Klein modes on $\Sigma$.  
Interestingly,  the absence of a noncommutative deformation 
is consistent with arguments 
that the small instanton singularity plays an important role in 
QCD dynamics \brodie.  Yang-Mills theories on non-commutative tori
do not have a small instanton singularity\Aharonyberk\nekrasov. 
For $\Sigma$ corresponding to a theory with a mass gap,  
the small instanton singularity exists precisely when 
$\int C = 2\pi {m\over N}$, with $m$ being
the 't Hooft flux.   If one instead considered a curve $\Sigma$ giving an
${\cal{N}}=2$  Super Yang-Mills theory,  a small instanton singularity     
would only exist if $\int C= 2\pi m$,  which is gauge equivalent to 
$\int C = 0$.  Furthermore,  for $\Sigma$ 
giving ${\cal{N}}=2$ Super Yang-Mills theory, one can decouple 
the Kaluza-Klein modes for all values of $\int C$.
At low energies,  we expect a continuous class of four dimensional 
Yang-Mills theories on a noncommutative torus.  
The difference between the case with a mass gap and the case without
is related to whether or not there is a $U(1)$ gauge field at low energies.  
Note that the non-commutative star product exists for $U(N)$ theories,  but not
for $SU(N)$.
 
The relation \reln\ is a very simple illustration of magnetic 
screening.  't Hooft has shown that in
an electrically confining $SU(N)/Z_N$ theory with a mass gap, 
magnetic fluxes must be light \thooft.  More precisely, the energy of
a magnetic flux must vanish exponentially with the area of the torus.
This is simply realized in the M-theory
construction.  Because of \reln, the 't Hooft flux is not the central
charge associated with a membrane.
In fact, the different magnetic fluxes correspond to classically degenerate 
M-fivebrane configurations.  This result
differs drastically
from what one would obtain by considering   
a compactification of the M-fivebrane giving ${\cal N}=2$ 
Super Yang-Mills theory.  
In that case the magnetic flux is related to an additive central charge,  
rather than an element of $Z_N$ labeling degenerate configurations.
 
We will also argue that $2\pi {m\over N}$ behaves 
like a two-form modulus of
the QCD string as well as the IIA string.  This means that the 
imaginary part of the action
of an MQCD string wrapping the two-torus $n$ times is given by 
$n\int_{T^3} C$.  
Again, we expect
this result to persist in the IIA limit,  so that the imaginary part of 
the action 
of a wrapped QCD string is given by $2\pi n{m\over N}$.  
In the large $N$ limit,  this becomes a continuous quantity.
Such a relation has been shown explicitly in two dimensions in the large
N limit \douglas\meone\metwo.
It was also conjectured to be true in four dimensions 
on the basis of properties of large $N$ QCD suggesting a T-duality 
invariance.  Such a duality would map,
\eqn\tdlty{\tau \rightarrow -{1\over\tau},}  
where $\tau$ is the K\"ahler modulus,
\eqn\kahl{\tau = {m\over N} + i{\Lambda^2 A \over 2\pi}.}
$A$ is the area of the torus, and $\Lambda^2$ is the QCD string 
tension.     

An argument due to 't Hooft states that in
an electrically confining $SU(N)/Z_N$ theory with a mass gap, 
magnetic fluxes must be light \thooft.  More precisely, the energy of
a magnetic flux must vanish exponentially with the area of the torus.
This is simply realized in the M-theory
construction.  Because of \reln, the 't Hooft flux is not the central
charge for a membrane.
In fact, when there is a mass gap, different 
magnetic fluxes correspond to classically degenerate M-fivebrane 
configurations.  This result
differs drastically
from what one would obtain by considering   
a compactification of the M-fivebrane giving ${\cal N}=2$ 
Super Yang-Mills theory.  
In that case the magnetic flux is related to an additive central charge,  
rather than an element of $Z_N$ labeling degenerate configurations. 

In the IIA limit, the M-fivebrane wrapped on $\Sigma$ becomes a 
configuration of N D4-branes 
suspended
between NS fivebranes.   For configurations giving ${\cal{N}}=2$ Super 
Yang-Mills on a commutative torus, 
the D4-brane worldvolume may also
contain integral numbers of fundamental strings, 
D2 branes and D0 branes \dougbranes.
However for configurations giving QCD-like theories with a mass gap, 
the allowed charges are more exotic.  There can be no D2 branes 
or fundamental IIA strings in the low 
energy theory.  This reflects the fact that both the magnetic and 
electric fluxes in QCD are elements of $Z_N$ rather than additive  
central charges.  However there may be Euclidean D0-branes with 
fractional charge.
When the theory is fully compactified on $\Sigma \times T^4$, the D0 
brane charge has a fractional 
part of the form 
${1\over N} m \wedge m$.
This naively appears to violate Dirac quantization in the presence of a 
single D6 brane.  However, due to the  nonzero $B = 2\pi {m\over N}$, 
additional charges are induced on the D6-brane which
preserve the Dirac quantization condition.

The organization of this paper is as follows.  In section II we briefly 
review the construction
of four dimensional gauge theories by wrapping M-fivebranes on calibrated 
two-cycles.  In
section III we perform an additional compactification on a two-torus,  
allowing a three form modulus in M-theory, and  show how this modulus
is related to the 't Hooft flux when the theory has a mass gap.
In section IV we show how the 't Hooft flux behaves like a two-form modulus for
the QCD string on a torus.  In section V we discuss the unusual D-brane 
charges allowed when there is a mass gap.

\newsec{M-fivebranes and QCD}

A variety of four dimensional Yang-Mills theories may be obtained 
by wrapping an M fivebrane on calibrated two cycles
embedded in $S^1 \times R^{10}$\edsolns\mqcdwitten.
Here we consider two-cycles corresponding to an SU(N) 
Yang-Mills theory with a mass gap. 
For instance a theory
in the same universality class as
${\cal{N}} = 1$ $SU(N)$ Yang-Mills is obtained by wrapping the 
five-brane on a particular holomorphic curve embedded in three
complex dimensions.  For details of this curve, the reader is referred to 
\mqcdwitten\mqcdooguri\mqcdstrassler.  
This theory becomes exactly ${\cal{N}} = 1, SU(N)$ Yang 
Mills in the weakly coupled IIA limit in which the radius of the 
$S^1$ goes to zero 
in eleven dimensional Planck units.
While taking this limit, the parameters of the curve are adjusted
to keep the QCD scale fixed.  At the same time one takes the limit 
$\Lambda_{QCD} l^{11}_p \rightarrow 0$ where $l^{11}_p$ is the 
eleven dimensional Planck
length.  We will begin by  studying the M theory limit in 
which the radius of the $S^1$ is large.  
However the quantities which we will compute are discrete and  
independent of the radius. 
For our purposes the degree of 
supersymmetry is unimportant so long as 
there is a
mass gap.  Our results are be equally applicable to the theory in
the same universality class as pure ${\cal{N}} = 0$ 
QCD \mqcdwitten\ which 
is obtained from a non-holomorphic but minimal area two cycle.

We will make use of two essential properties of two-cycles $\Sigma$ 
corresponding to $SU(N)$ Yang-Mills theories with a mass gap.
The first property is that $H_1(\Sigma, Z)$ is generated by 
a cycle $S^{1\prime}$ wrapping the $S^1$ of 
space-time N times \mqcdwitten\mqcdooguri.  
The second property is that there are no square 
integrable harmonic one-forms 
on $\Sigma$. A compactification of $\Sigma$ 
obtained by adding points at infinity
gives a curve of genus zero \mqcdwitten.  
Because of the latter property,  there are no 
light states after dimensional 
reduction.  At low energies,  the only solution of the equations of 
motion for
the self dual three-form field strength of the M5-brane is 
\eqn\key{T^{(3)} = db^{(2)} - C^{(3)}|_{M5} = 0,}
where $b^{(2)}$ is the two-form gauge potential of the M-fivebrane,
and $C^{(3)}|_{M5}$ is the pullback of the bulk 
three-form potential to the M-fivebrane.
Note that if we instead considered a curve 
$\Sigma$ giving ${\cal{N}} = 2$ super Yang-Mills
theory,  then there would be normalizable harmonic 
one forms associated with 
massless $U(1)$ vector multiplets in the low 
energy theory \edsolns. 
If the M-fivebrane worldvolume $X$ has a  non-trivial 
$H_3(X,Z)$,  there may be several 
vacua corresponding to solutions of \key.  
We assume that the background
$C^{(3)}$ is flat, so $dC^{(3)} = 0$.  We will see that these 
vacua are discrete.

\newsec{Three-form moduli and 't Hooft fluxes}

Let us now compactify M theory on $S^1 \times T^2 \times R^8$,  
and wrap an
M-fivebrane on $\Sigma \times T^2 \times R^2$.  The low energy theory is 
an SU(N) Yang-Mills theory on $T^2 \times R^2$ with a mass gap.
The M-fivebrane worldvolume now has non-trivial three cycle 
$S^{1 \prime} \times T^2$.  Since the M5 brane theory contains 
strings coupling to $b^{(2)}$,  there is a flux quantization condition
\eqn\fluxq{\int_{S^{1 \prime} \times T^2} db^{(2)} = 2\pi m,}
where $m$ is an integer.   Then because $T^{(3)} = 0$,  
\eqn\cquant{\int_{S^{1 \prime} \times T^2} C^{(3)}|_{M5} = 2\pi m.}
Since $S^{1 \prime}$ wraps N times around $S^1$,  the 
bulk three-form modulus is
given by 
\eqn\cqanttwo{\int_{S^1 \times T^2} C^{(3)} = 2\pi {m\over N}.}

In order to get solutions for other values of $\int C^{(3)}$,
one must expand $T^{(3)}$ in the Kaluza Klein modes on $\Sigma$.
Note that in general $\Sigma$ will change in the presence of   
a non-zero $T^3$,  however the property that there are no normalizable
harmonic one-forms persists.
In this case we do not expect varying $\int C^{(3)}$ to give a continuous 
class of four dimensional Yang-Mill theories on non-commutative tori. 

We will now argue that the the integer $m$ is 
the 't Hooft flux. 
To do so we will work in the IIA limit.  In this limit the M5-brane 
becomes a set of $N$ coincident D4-branes stretched between 
NS5-branes.  In the absence of the $NS5$-branes,  the low 
energy theory would
have a $U(N)$ gauge group.  However in our case there is a mass gap, 
and the theory is $SU(N)$.  The $U(1)$ degree of freedom is 
frozen: ${1\over N} tr F-B_{NS} = 0$.  This follows from
dimensional reduction of $T^{(3)} = 0$.  
If the D4-brane is compactified
on a torus,  then fields on the torus are periodic up to gauge 
transformations
$U^1$ and $U^2$.  These gauge transformations 
are subject to a consistency
condition \zram,
\eqn\cons{U^1(x^1,x^2)U^2(x^1+ 2\pi R^1,x^2)
	U^{1\dagger}(x^1+ 2\pi R^1 ,x^2 + 2\pi R^2)U^{2\dagger}
	(x^1,x^2 + 2\pi R^2) = I,}
This condition permits the existence of fundamental matter,  which 
appears when strings
end on the D4 branes.  Note that \fluxq\ follows 
from an analogous condition.
The $U$'s may be broken up into a $U(1)$ factor and
an $SU(N)$ factor,  so that the above condition becomes
\eqn\consfac{e^{i\int_{T^2} {1\over N} Tr F}e^{2\pi i {m\over N}} = I,}
where $m$ is the $SU(N)$ 't Hooft flux.   
Since ${1\over N} tr F = B_{NS}$,
we find the 't Hooft flux is precisely the discrete quantum number 
discovered in the M-theory approach.  
The fact that the 't Hooft flux is only defined
modulo $N$ is reflected in the fact that $2\pi$ shifts in $\int B_{NS}$ 
are gauge transformations. 
Thus the 't Hooft fluxes are associated with 
degenerate classical vacua of
the M5-brane, all of which have $T^{(3)} = 0$.  
This demonstrates magnetic screening and 
is precisely what one expects for
for an electrically confining theory with a mass gap \thooft.

The constraint $F_{U(1)}-B = 0$ arising from the existence of 
a mass gap means that the theory is a standard gauge theory on
a commutative torus.  In \jabone\jabtwo\peimingho\ this quantity 
was argued to be a 
local measure of non-commutativity on the brane worldvolume. 
A simple way to see that the small instanton singularity is not
resolved is to perform a T-duality along one of the cycles of
the torus.  The $B$ field then becomes the real part of the complex
structure of the dual torus.  The $N$ D4-branes become a D3-brane at an
angle determined by ${1\over N}TrF$.  Consider wrapping a Euclidean 
D1-brane on the cycle of the torus on which the T-duality is 
performed.  Because of the relation $B=2\pi {m \over N}$ relating
the real part of the complex structure to the angle of the D3-branes,
the D1-brane is at a 90 degree angle to the D3-brane,  giving a 
BPS configuration.  There is a Coulomb branch describing motion
relative to the D3-branes.  The D1 can be placed such that it intersects
the D3-branes.  Then upon undoing the T-duality,  the D1-brane becomes a 
Euclidean D0-brane corresponding to a small instanton.
It has been argued \brodie\ that the small instanton singularity plays an 
important role in  QCD dynamics.  Our result supports this claim,
since a non-commutative deformation does not exist when there is a mass gap.
Note that this is related to the freezing of the $U(1)$ degree of freedom.
A non-commutative star product does not exist when this degree of freedom
is frozen.  In $U(N)$ gauge theories on a non-commutative torus,  gauge 
transformations mix the $SU(N)$ and $U(1)$ components.

It is interesting to note that in the $N\rightarrow\infty$ limit of a 
of the theory with with a mass gap,  one is allowed a 
continuous range of values for $B$,  all of which correspond to
a gauge theory on a commutative torus.

\newsec{$B$ field for the QCD string}

It has long been suspected that large N QCD is a string theory.
We now wish to prove the conjecture \meone\metwo\ that the 't Hooft flux
behaves like the real part of a two-form modulus of the QCD-string. 
To do so we will show that the 
Euclidean action
of a QCD string wrapping $T^2$ $n$ times has 
an imaginary part given by
$2\pi n {m\over N}$  Let us first compute the 
imaginary part of the action of
a wrapped MQCD string. 
The MQCD string \mqcdwitten\  is an open membrane 
ending on the M5-brane.  The worlvolume of the MQCD string is 
$I \times {\cal{M}}$,  where ${\cal{M}}$ is 
the world sheet of the QCD string and $I$ is an open interval generating
the relative homology $H_1(Y/\Sigma,Z_N)$. 
$Y$ is the space in which $\Sigma$ is embedded. 
For a detailed discussion of the MQCD string,  the reader is referred to
\mqcdwitten\mqcdstrassler.  The action of the 
open M-twobrane has an imaginary piece: 
\eqn\mtwo{Im S= \int_{I\times {\cal{M}}} C^{(3)}|_{M2} - 
		\int_{\del I \times {\cal{M}} } b^{(2)},}
where $C^{(3)}|_{M2}$ is the pullback of the 
bulk three-form to the membrane,
and $b^{(2)}$ is the two-form of the M-fivebrane world volume.
The second term is necessary for gauge invariance of the 
open
membrane action.  Under $C^{(3)}\rightarrow  C^{(3)} + d \lambda^{(2)}$,
$b^{(2)}$ transforms to cancel the the 
non-invariance due to the boundary:
$b^{(2)} \rightarrow b^{(2)} + \lambda^{(2)}|_{M2}$.
Because $db^{(2)} - C^{(3)}|_{M5} = 0$,
we may rewrite \mtwo\ as
\eqn\mtwotwo{Im S = \int_{(I - \Omega) \times {\cal{M}}} C^{(3)},}
where $\Omega$ is a chain in the M5-brane with boundary 
$\del I \times {\cal{M}}$.  The MQCD string is homotopic to the IIA 
string, which is a membrane with the world volume
$S^1 \times {\cal{M}}$ \mqcdwitten.  Therefore the above
expression becomes  
\eqn\mtwo{Im S = \int_{S^1 \times  {\cal{M}}} C^{(3)}.}
So if ${\cal{M}}$ wraps $T^2$ $n$ times,  
\eqn\rap{Im S = 2\pi n{m\over N}.}
This result is discrete and independent of the radius 
of the $S^1$, so we expect it
to hold in the IIA limit. 
In the large $N$ limit $2\pi m\over N$ becomes continuous,  and behaves 
as the two-form modulus for the QCD-string,  as well as the IIA string.

\newsec{D-brane charges in theories with a mass gap}

The allowed D-brane charges in theories with a mass gap differ 
substantially from those without a mass gap.  
If we had  considered a brane construction of an  
${\cal{N}} = 2$ theory instead,  the D2 brane charge and the 
fundamental string charge would equal the magnetic and
electric fluxes respectively.  However   
the fluxes in QCD are very different objects. 
The term in the M5-brane action 
of the form $S = \int C^{(3)} \wedge db^{(2)}$ vanishes in
MQCD because 
$db^{(2)} - C^{(3)} = 0$.  Due to the
self duality of $db^{(2)} - C^{(3)} = 0$, this term
becomes  
\eqn\temr{S = \int C^{(3)}_{RR} \wedge Tr (F-B) + 
\int B_{NS} \wedge
Tr^*(F-B),} 
in the IIA limit.
This vanishes, so there are no
states in the low energy theory carrying  either D2 brane
or fundamental string charge.  

In brane constructions of QCD, the Euclidean 
D0-brane charge may be fractional.  The 
importance of such fractional charges for confinement 
and dynamical generation of  
superpotentials has been discussed in \brodie.
We will show that the such fractional charges are 
consistent with Dirac quantization
in the presence of a D6-brane.   
The Chern-Simons term in the D4 brane action which 
determines whether there
is also D0-brane charge is given by \dougbranes
\eqn\zerobrane{\int C^{(1)}_{RR} \wedge 
Tr \left[ (F-B) \wedge (F-B)\right]}
Since $Tr (F-B) = 0$,  the D0-brane charge 
is given by the $SU(N)/Z_N$ contribution
to the instanton number,  which may be fractional \thooftor. 
If one compactifies $SU(N)/Z_N$ QCD on $T^4$,  
then the instanton number is 
given by 
\eqn\fracinst{ {1\over 16\pi^2} \int {\cal F} \wedge {\cal F} = 
\nu + {m \wedge m \over N},}
where $\nu$ is an integer and $m$ are the integer 't Hooft 
fluxes.
The fractional piece apparently violates Dirac Quantization in 
the presence of a single
D6-brane,  since $Q_6Q^{\prime}_0$ is not an integer multiple of $2\pi$.
If a D6-brane is present then the theory on the D4 brane 
contains massive fundamental matter due to string stretched between the 
D6-brane and the D4-brane.  Since fundamentals  of $SU(N)$
are charged under the $Z_N$ center,  it would naively seem that 
't Hooft flux must vanish modulo $N$ and the instanton number must be 
integer. 
However the matter introduced by the D6-brane is in a fundamental 
representation of $U(N)$,  and a nontrivial $SU(N)$ 't Hooft twist
may be cancelled by a $U(1)$ twist, as in \cons.  The 
introduction of fundamental matter by a 
D6-brane does not prohibit non-trivial 
't Hooft fluxes or fractional D0-brane charge.  

The Dirac quantization problem is resolved as follows.
When the 't Hooft flux is non-zero,  there is 
also a background $\int B$ given
by the relation \IIAlimit.  For non-zero $\int B$ a D6-brane 
carries other D-brane charges \dougbranes.
The non-integer terms which threaten to violate
Dirac quantization turn out to cancel just as they 
would for a pair of dyons in the presence of a non-zero theta parameter.
Suppose that the $D4$ brane on which the 
Yang-Mills theory lives is extended in the 
directions $01234$, with the directions $0123$
compactified on $T^4$.  The D4-brane is bounded 
by NS-5 branes at fixed values of $x^4$.
Consider the action of a D6-brane with the 
world-volume $T^4 \times S^3$, where $S^3$ 
is embedded in
the directions $56789$.  Also, suppose that 
the D6-brane is at a value of $x^4$ between the 
NS5-branes and does not 
intersect the D4-D0 system.  The path integral for the D6 brane
contains a term
\eqn\dq{e^{i\left(\int_{T^4 \times S^3} C_{RR}^{(7)} - 
	{B \wedge B \over 4\pi^2} \int_{S^3} C_{RR}^{(3)}\right)},}
The second term reflects the fact that in the presence of
$B$,  a D2-brane charge is induced on the D6-brane.  
A D4-brane charge is induced as  
well, however
the corresponding term in the action does 
not effect Dirac quantization.  This is 
because the the D4-brane is hodge dual to a 
D2-brane, and the D2-brane charge vanishes on the 
D4-brane associated with the 
Yang-Mills theory.
Because of the D4-D0 system, the Ramond-Ramond 
potentials $C_{RR}^{(7)}$ and 
$C_{RR}^{(3)}$ are not globally
defined.  For \dq\ to be well defined,  one must have
\eqn\wrks{e^{i\left(\int_{T^4 \times S^4} dC_{RR}^{(7)} - 
	{B \wedge B\over 4\pi^2} \int_{S^4} dC_{RR}^{(3)}\right)} = 1,}
where $S^4$ is embedded in the directions $56789$ 
and surrounds the D4-D0 system, which is 
a point in $56789$.
$\int dC^{(7)}$ is simply the D0-brane charge,  
with fractional piece $m \wedge m \over N$.
Similarly $\int dC^{(3)}$ is the D4-brane charge, or N.  Since 
$B=2\pi{m\over N}$, equation \wrks\ is satisfied.   
Roughly speaking, we have found that  
\eqn\chges{Q^{(0)} \wedge  Q^{\prime (6)} + 
	Q^{(2)} \wedge Q^{\prime (4)} - Q^{(4)} \wedge Q^{ \prime (2)}=
	2 \pi l,}
where the $Q^{\prime}$ charges are associated with
the D6-brane,  and the $Q$ charges are associated with the D4-branes.

\bigbreak\bigskip\bigskip

\centerline{\bf Acknowledgments}\nobreak

This work was supported in part by DOE under contract 
No. DE-AC02-76-ER-03071.
I am especially grateful to Burt Ovrut and Edward Witten for
enlightening conversations. 
I also profited greatly from discussions with 
Edna Cheung, Miriam Cvetic, Ori Ganor,  Antal Jevicki, 
Robert de Mello Koch,
Morten Krogh, Sanjaye Ramgoolam, and Daniel Waldram

\listrefs

\end